\documentclass[12pt]{article}
\usepackage{geometry}
\usepackage{hyperref}
\usepackage{mcite}
\usepackage{mathrsfs}
\usepackage{amssymb,amstext,amsmath,amsthm,eucal}

\newcommand{\hodge}{\ast}
\newcommand{\Cl}{\mathrm{C}\ell}
\newcommand{\RR}{\mathbb{R}}
\newcommand{\CC}{\mathbb{C}}
\newcommand{\AdS}{\textrm{AdS}}

\newcommand{\Mat}{\mathrm{Mat}}

\newcommand{\eps}{\varepsilon}
\newcommand{\dvol}{\mathrm{dvol}}
\newcommand{\spin}{\mathfrak{spin}}
\newcommand{\su}{\mathfrak{su}}
\newcommand{\lie}{\mathcal{L}}
\newcommand{\grpel}[1]{\mathcal{#1}}
\newcommand{\liel}[1]{\mathbf{#1}}

\newcommand{\Tr}{\mathrm{Tr}}

\newcommand{\rbs}[2]{\left({#1}\,|\,#2\right)}
\newcommand{\magn}{h}

\begin{document}

\title{The NUT in the N=2 Superalgebra}
\author{George Moutsopoulos
\\ [.2in] Department of Physics  and\\  State Key Laboratory of
Nuclear Physics and Technology,\\ Peking University, Beijing 100871,
China\\[.2in]\small\texttt{gmoutso@googlemail.com}\\[.1in]{}}
\date{}

\maketitle

\begin{abstract}
 We study how the supersymmetry algebra extension of $\mathcal{N}=2$ supergravity with central charges describes the NUT charge in a duality-covariant way. The stationary BPS states of $\mathcal{N}=2$ supergravity can be embedded in the flat space solution of the timelike reduced theory. Upon reduction, the Killing superalgebra extension and the global superalgebra of charges are described covariantly under the coset structure of the three-dimensional theory. The integral of the Nester-Witten form given by Gibbons and Hull for $\mathcal{N}=2$ has a covariant generalization that includes the NUT charge.
\end{abstract}

\section{Introduction}
A lot of work in high energy physics focuses on supersymmetry. In supersymmetric field theories the Poincar\'e algebra of symmetries is extended with spinorial Grassmann-odd symmetries, so that the bracket of two odd elements closes on the translations. Information on the theories' solitonic solutions can be extracted from extending the Poincar\'e superalgebra with additional even elements. The presence of these elements corresponds to turning on the charges of the embedded solitons, see e.g.~\cite{Witten:1978mh}. A remarkable result of this is the prediction of the fundamental M-branes in the elusive M-theory from the M-theory superalgebra alone~\cite{Townsend:1997wg,Bergshoeff:1997bh}. When a theory possesses dualities, as with the case of S-duality in type IIB string theory, the maximal extension of the Poincar\'e superalgebra is covariant under the duality. Upon dimensional reduction the maximal extension should be described covariantly under the larger U-duality of the theory.

Supergravity theories have a local version of the Poincar\'e symmetry superalgebra. The fields of these theories transform under supersymmetry with spinor parameters that depend on the spacetime coordinates. The commutator of two supersymmetry transformations closes on the even symmetries of the theory, that is, diffeomorphisms, local Lorentz transformations and gauge shifts. The finite-dimensional Killing superalgebra of a background contains the supersymmetry transformations and the vector fields that leave the background invariant, see e.g.~\cite{FigueroaO'Farrill:2004mx}. The Poincar\'e superalgebra can then be recovered as the Killing superalgebra of the flat space solution, a maximally supersymmetric vacuum of the underlying quantum theory.

The Poincar\'e superalgebra can be extended maximally, in the sense of~\cite{FigueroaO'Farrill:2008ka}, for trivial geometric reasons. It is natural to ask if other backgrounds similarly admit an extension of their Killing superalgebra. It has been shown that a maximal extension is also possible for the Freund-Rubin (${\AdS}_m\times S^n$) backgrounds, but the maximally supersymmetric pp-wave does not admit a maximal extension~\cite{FigueroaO'Farrill:2008ka}. It remains an open question how to prescribe an extension of the Killing superalgebra of the latter, or any background, that is also physically meaningful. If such an extension exists, it would provide an invariant of the background that distinguishes it from those backgrounds that have the same Killing superalgebra.

There is a nice connection between the Killing superalgebra of an asymptotically flat background and the superalgebra of asymptotic symmetries~\cite{Hull:1983ap}. If a background has a flat limit, then its Killing superalgebra should limit to a subalgebra of the Poincar\'e. It is even more useful to consider their extended version. We can define the Killing superalgebra extension of an asymptotically flat background to be the subalgebra of the Poincar\'e extension that is generated from its Killing spinors, as a subspace of the constant spinors in the flat limit. Using a generalized Nester-Witten form~\cite{Witten:1981mf,*Nester:1982tr}, one can prove a Bogomolny bound that the charges of the theory should satisfy and show how the existence of a Killing spinor implies the bound is saturated~\cite{Gibbons:1982fy}.

In~\cite{Hull:1998mh}, Hull argued that the dimensional reduction of the charges of M-theory can be dualized to describe NUT charge. The NUT charge can be calculated using the integral of the Nester-Witten form
\[ \int_{S^2_\infty} \overline{\eps}\,{\Gamma_{MN}{}^P\,\nabla_P\eps}\,\hodge\,dX^M\wedge dX^N~,\]
where the indices are summed over $M,N,P=0,1,\dots 10$, $S^2_\infty$ is the asymptotic boundary at ``spatial infinity'' and $\eps$ is a Majoranna spinor field that limits at spatial infinity to a constant spinor. The integral gives the ADM mass plus a term $\int \omega_{3}\wedge\xi_{6}$, where $\omega_3$ is the three-form defined at asymptotic infinity via the spin connection and $\xi_6$ is a six-form from the spinor bilinear. The latter term is related to the euclidean NUT charge that is carried by Kaluza-Klein monopoles~\cite{Sorkin:1983ns,Gross:1983hb}. Upon dimensional reduction, the charges of the theory transform into each other under U-duality.

In this paper we are interested in the lorentzian NUT charge~\cite{Bossard:2008sw} in four dimensions for asymptotically Misner-flat spacetimes~\cite{Misner:1963fr}. Given an asymptotic, $r\rightarrow\infty$, timelike Killing vector $K$, the Komar mass $M$, the magnetic charge $H$ and the electric charge $Q$ are defined as
\begin{subequations}
 \begin{align}\label{eq:Komardef1}
 M&=\frac{1}{2}\int_{S_\infty^2}\hodge d K^\flat~, &
 H&=\int_{S^2} F~,&
 Q&=\int_{S^2} \hodge F ~,
\end{align}
where $F$ is the Maxwell field-strength, $K^\flat=g(K,-)$ is the metric dual of the vector field $K$, and $S^2$ is a two-sphere that is large enough to capture any internal sources. We accordingly define the NUT charge as a dual to the Komar mass
\begin{equation}\label{eq:Komardef4}
 N = \frac{1}{2}\int_{S_\infty^2} d K^\flat~.
\end{equation}\end{subequations}
For a non-zero $N$, this formula makes sense if the isometry $K$ defines a non-trivial fibration in a region $M$ close to $r\rightarrow\infty$, where $K$ acts free and proper. The two-spheres $S^2$ and the limit $S^2_\infty$ are then two-spheres in the space of orbits $\tilde{M}$ of the bundle $M\stackrel{\pi}{\rightarrow}\tilde{M}$ and the quantities integrated are the pullbacks of the forms we wrote. Since only circle fibrations can be non-trivial, $M$ necessarily has closed timelike curves. We will work in the region $M$ or $\tilde{M}$, but our results are unambiguous as an invariant of the whole background.

A motivation for our work stems from the results of~\cite{Argurio:2008zt}, see also \cite{Argurio:2008nb}. By complexifying the Nester-Witten form
\[ \mathrm{Re}\left( \overline{\eps}\,\Gamma_5 \,\Gamma_\mu \,\nabla_\nu\eps \right)dx^\mu\wedge dx^\nu
\longrightarrow
 \overline{\eps}\,\Gamma_5 \,\Gamma_\mu \,\nabla_\nu\eps \,dx^\mu\wedge dx^\nu~,\]
where $\mu,\nu=0,1,2,3$ and $\eps$ is an asymptotically constant Dirac spinor field, $\eps\rightarrow\eps_0$, the authors showed that its integral over asymptotic infinity includes the term
\[ N_{\underline{\mu}} \, {\overline{\eps}_0}{\Gamma^{\underline{\mu}}\Gamma_5\eps_0}~ \]
that matches the NUT charge: $N_0=N$. This is accomplished by choosing a ``triangular gauge'' for the vielbein. However, as they point out, the spinor bilinear is not symmetric, or else the supercharges cannot be considered hermitian. The NUT space has a topology $S^3\times\RR^+$, which is very different to Minkowski's, so it is hard to interpret $N^{\underline{\mu}}$ as a four-vector. It is natural to interpret this expression in the three-dimensional $\tilde{M}$ and we will do so in section \S\ref{sec:SusyNW}. 

We will focus on $\mathscr{N}=2$ pure supergravity, the supersymmetric extension of Einstein-Maxwell theory in $3+1$ dimensions. This is the simplest theory with the desired features that we are after. We would like to investigate how an extension of the Killing superalgebra can accommodate for the NUT charge. By this we mean that the extension distinguishes between backgrounds of different $(M,N,Q,H)$ charges. 
 For an asymptotically flat space, the electromagnetically extended $\mathscr{N}=2$ Poincar\'e superalgebra of charges is
\begin{equation}\label{eq:IntroPoinc}
 \{ Q_{\alpha}^\mathcal{I},Q_\beta^\mathcal{J}\}=(\Gamma^\mu \,C^{-1})_{\alpha\beta}\delta^{\mathcal{I}\mathcal{J}} P_\mu + \epsilon^{\mathcal{I}\mathcal{J}}(C^{-1})_{\alpha\beta}Q +  \epsilon^{\mathcal{I}\mathcal{J}}(\Gamma_5\,C^{-1})_{\alpha\beta}H~.
\end{equation}
We will argue that, in contrast to the inclusion of $N_{\underline{\mu}}$ in the superalgebra of \cite{Argurio:2008zt}, this contains all the information we need, including the NUT charge. However, a NUT-charged spacetime has closed timelike curves and an ADM definition of the charges is subtle, see e.g.~\cite{Hunter:1998qe}. For this reason, it is more natural to use Komar's definition and describe the extended superalgebra of charges in the timelike reduced theory. We will continue to talk about the ``extension of the Poincar\'e superalgebra'', although we are not including the boost elements of the Lorentz algebra.

Given a timelike isometry, Einstein-Maxwell theory reduces to a sigma model on the coset space $SU(1,2)/U(1,1)$ coupled to euclidean gravity~\cite{Kinn:Gen1973}. In section \S\ref{sec:StatFerm} we find how the three-dimensional gaugino, gravitino and Killing spinors transform under $U(1,1)$. We do this by reducing the supersymmertry variation of the four-dimensional gravitino. With this knowledge, in section \S\ref{sec:SusyTriplet}, we show that the extended Killing superalgebra in three dimensions has a natural description under the coset structure. The odd-odd bracket defined on a Killing spinor gives a three-dimensional Killing vector and a triplet of scalars transforming in the adjoint of $\su(1,1)$ that is covariantly constant. These even elements originate from the reduction of the four-dimensional Killing vector and the two scalars from the four-dimensional Killing spinor bilinear. The latter two scalars transform under the electromagnetic $\mathfrak{u}(1)$ symmetry in four dimensions, but in three dimensions they are components of the scalar triplet that transforms under $\su(1,1)$. The NUT-charged spacetimes are now globally flat at asymptotic infinity (in $3+0$ dimensions) and the sigma model limits to the trivial map. The extended Killing superalgebra of the NUT-charged BPS backgrounds are embedded like this in the extended Killing superalgebra of flat spacetime.

Finally, in section \S\ref{sec:SusyNW} we covariantize the Nester-Witten form under the coset stucture. That is, we covariantize \eqref{eq:IntroPoinc} under $\mathfrak{u}(1,1)$ so that the NUT charge $N$ appears on an equal footing to the mass $M$ and the charges $Q$ and $H$. In this form the odd-odd bracket of supercharges remains symmetric. Additionally, we can show why the term $N_{\underline{0}}$ coming from the complexified Nester-Witten form gives the NUT charge. In our formalism we avoid an ill-defined integral over spatial infinity that was in four dimensions. We find that the term $N_{\underline{0}}$ belongs to a \emph{different} representation of $\mathfrak{u}(1,1)$ than the one in which the covariantized Nester-Witten expression belongs to.

We begin with a description of $\mathscr{N}=2$ supergravity and develop our notation. In section~\ref{sec:StatBos} we describe the coset space parametrized by the bosons. This is a necessary step as the fermions live on bundles defined by the bosonic sector. By using the four-dimensional supersymmetry transformation of the gravitino, we derive in section~\ref{sec:StatFerm} the representation of the three-dimensional fermions. Then, in section~\ref{sec:StatBPS}, we give the stationary BPS solutions of the theory. In section~\ref{sec:Susy} we begin with a field-theoretic motivation of the Killing superalgebra extension of $\mathscr{N}=2$ in four dimensions. We then show in section \ref{sec:SusyTriplet} how this reduces to a $U(1,1)$-covariant structure in three dimensions. As mentioned, the three-dimensional Killing spinors square to a three-dimensional Killing vector and a gauge-covariant constant scalar in $\su(1,1)$. In section \ref{sec:SusyBrack} we describe the rest of the Killing brackets. Finally, in section~\ref{sec:SusyNW} we discuss the supersymmetry algebra of charges and the Nester-Witten form in the reduced theory. There we show that the gauge-covariant $\su(1,1)$ scalar from the square of Killing spinors leaves the charges of the theory, in the fundamental of $\mathfrak{u}(1,1)$, fixed. Appendix~\ref{app:su12} contains details on the representation we used in section~\ref{sec:StatBos} and appendix~\ref{app:D} has some details on the BPS Killing spinor equation of section~\ref{sec:StatBPS}.

\section{Stationary solutions}\label{sec:Stationary}
The field content of pure $\mathscr{N}=2$ supergravity is the vielbein $e^{\underline{\nu}}_\mu$, a Maxwell gauge field $A_\mu$ and a complex-valued gravitino $\psi_\mu$~\cite{Ferrara:1976fu}. We use a mostly minus signature $(+,-,-,-)$ and the Clifford representation $\{\Gamma^{\underline{\mu}},\Gamma^{\underline{\nu}}\}=-2\eta^{\underline{\mu}\,\underline{\nu}}$, which is real and isomorphic to $\Mat(4,\RR)$. The gravitino transforms under a supersymmetry parameter $\eps$ as
\begin{equation}\label{eq:deltapsi4}
 \delta\psi_\mu=D_\mu\eps:=\left(\nabla_\mu-\frac{i}2 F \,\Gamma_\mu\right)\eps+O(\psi^2)~,
\end{equation}
where the field-strength $F=dA$ acts on the Dirac spinor field $\eps$ through Clifford multiplication. With these conventions, the bosonic equations of motion can be found by varying the gravitino equation of motion, $\Gamma^{\mu\nu\rho}D_\nu\psi_\rho=0$, under an arbitrary supersymmetry parameter. This is equivalent to $\Gamma^\mu [D_\mu,D_\nu]=0$, from which we extract the Einstein-Maxwell equations
\begin{equation}\label{eq:EMeom}
\begin{aligned}
 \frac{1}{2}R_{\mu\nu}&=F_{\mu\sigma}F^\sigma{}_\nu+\frac{1}{4}F^{\rho\sigma}F_{\rho\sigma}\eta_{\mu\nu}\\
dF&=d\hodge F=0~.
\end{aligned}
\end{equation}

The spinors of $\mathscr{N}=2$ supergravity are complex (Dirac) spinors. The Lorentz invariant antihermitian product $\overline{\eps_1}\eps_2$ of two such spinors $\eps_1$ and $\eps_2$ is a complex scalar. It is convenient to use the notation $\rbs{\eps_1}{\eps_2}$ for the real part of the inner product, which is symplectic on Dirac spinors. We are interested in real superalgebras so we prefer to use the symplectic inner product from the onset. One-forms are skew-symmetric with respect to the symplectic inner product and the symmetric spinor bilinears that one can make are: $\rbs{\eps}{i\,\eps}$, $\rbs{\eps}{\Gamma_\mu\,\eps}$, $\rbs{\eps}{\Gamma_{\mu\nu}\,\eps}$, $\rbs{\eps}{i\,\Gamma_5\Gamma_\mu\,\eps}$  and $\rbs{\eps}{i\,\Gamma_5\,\eps}$. In index-free notation, the supersymmetry variations of the vielbein and Maxwell field are
\begin{align*}
\delta e^{\underline{\mu}}&=\rbs{\eps}{\Gamma^{\underline{\mu}}\psi}\\ 
\delta A&=-\rbs{\eps}{i\,\psi}
 ~.
\end{align*}
A lot is known about this theory and its supersymmetric configurations have been classified in~\cite{Tod:1983pm,Tod:1995jf}.

If a bosonic solution has a timelike isometry $K=\partial_t$, we can write the metric $g$ and $A$ in an adapted frame
\begin{equation}\label{eq:stansatz}\begin{aligned}
 g&=V\,(dt+B)^2+V^{-1}\,\tilde{g}\\
 A&=-\phi\,(dt+B)+\tilde{A}~,
\end{aligned}\end{equation}
so that $V$, $B$, $\tilde{g}$, $\phi$ and $\tilde{A}$ do not depend on time and have components only in the other three directions. Note that $V$, $\theta :=dt+B$, $\tilde{g}$, $F=dA$ and $d\phi=\iota_K F$ are globally defined on the fiber bundle whose fibers are the orbits of the isometry, whereas $B$ and $\tilde{A}$ are gauge fields that are defined locally and up to gauge transformations.

A lagrangian that gives the Einstein-Maxwell equations is
\[\mathscr{L}=R(g)\,\dvol_g+2\,F\wedge{\hodge} F~,\]
which, after formally integrating time, becomes
\begin{multline*}R(\tilde{g})\,\dvol_{\tilde{g}}-\frac{1}{2} V^2 dB\wedge{\tilde\hodge} dB-\frac{1}{2V^2} dV\wedge{\tilde\hodge} dV 
\\+\frac{2}{V}d\phi\wedge{\tilde\hodge} d\phi+2 V \left(d\tilde{A}-\phi \, dB\right)\wedge{\tilde\hodge}\left(d\tilde{A}-\phi \, dB\right)
~,\end{multline*}
where $\tilde{\hodge}$ is the three-dimensional Hodge operator. 
The equation of motion for $\tilde{A}$ is $d\,{\tilde\hodge} V(d\tilde{A}-\phi \, dB)=0$ and for $B$ is $d\,{\tilde\hodge}V^2dB=-2\,d\phi\wedge V\,\tilde\hodge(d\tilde{A}-\phi\,dB)$. We can dualize $B$ and $\tilde{A}$ into the scalars $\omega_0$ and $\magn$ by adding the Lagrange multipliers $+d(\omega_0+2\phi  \magn)\wedge dB$ and $-{4}\,d\magn \wedge d\tilde{A}$~\cite{Maison:1984tg}. Integrating the field-strengths $dB$ and $d\tilde{A}$ gives the dual lagrangian
\begin{equation}\label{eq:3dLag}
\tilde{\mathscr{L}}= R(\tilde{g})\,\dvol_{\tilde{g}}-\frac{2}{V^2} \omega_1\wedge{\tilde\hodge} \omega_1-\frac{1}{2 V^2} dV\wedge{\tilde\hodge} dV 
+\frac{2}{V}d\phi\wedge{\tilde\hodge} d\phi+\frac{2}{V} d\magn\wedge{\tilde\hodge} d\magn~,
\end{equation}
where
\[\omega_1=\frac{1}{2}d\omega_0-\phi \, d\magn +  \magn \, d\phi.\]
We could have equivalently reduced the Einstein-Maxwell equations of motion directly, as in~\cite{Heusler:1997am}. The one-form $\omega_1$ is the twist one-form
\[\omega_1=\frac{1}{2}{\hodge} (K^\flat \wedge dK^\flat)~,\]
where $K^\flat=g(-,K)$ is the metric dual of $K$, and the field-strength is
\[F=\theta\wedge d\phi-{\hodge}(\theta\wedge d\magn)~.\]

We find the relation between the four- and three-dimensional hodge operation useful. For any $n$-form $\tilde{b}_n$ with legs in the three-dimensional manifold $\tilde{M}$, we have  
\begin{align*}
{\hodge}(\theta\wedge \tilde{b}_n)=V^{n-2}\,{\tilde\hodge} \tilde{b}_n\\\intertext{and}
{\hodge} \tilde{b}_n =-V^{n-1}  \,{\tilde\hodge} \tilde{b}_n\wedge\theta~.
\end{align*}
It is also convenient to use a Clifford algebra relation for the hodge dual of an n-form $b_n$
 defined in four dimensions. It is $\hodge b_n = (-1)^{\frac{n}{2}(n-1)}\Gamma_5 \, b_n$, where $\Gamma_5$  is the volume form acting through Clifford multiplication. Likewise, for an $n$-form $\tilde{b}_n$ in three dimensions it is $\tilde\hodge \tilde{b}_n = (-1)^{\frac{n}{2}(n-1)}\gamma_5 \, \tilde{b}_n$, where $\gamma_5$ is the three-dimensional volume form. We use these relations, in particular, to reduce the four-dimensional gravitino variation.

We observe that for stationary solutions, electromagnetic duality becomes a manifest symmetry at the level of the effective lagrangian. This is realized by a rotation of $\phi+i\,\magn$. This $U(1)$ is further enlarged to the ``hidden'' symmetry group $SU(1,2)$, where the scalars $\phi$, $\magn$, $V$ and $\omega_0$ parametrize the coset space $SU(1,2)/U(1,1)$. We will first describe the scalars under the coset structure and then the reduced gravitino and the Killing spinor equation. Finally, we will present the elementary BPS solutions.

\subsection{Coset Space}\label{sec:StatBos}
The standard description of a symmetric space sigma model on $G/H$ chooses a coset representative $\grpel{V}(x)\in G$ in the equivalence class $\grpel{V} \equiv \grpel{}{h}(x)\,\grpel{V}$ for any $\grpel{}{h}\in H$. The subgroup $H$ is fixed by a Cartan involution $\sigma$ that splits the Lie algebra of $G$ into $\mathfrak{g}=\mathfrak{k}\oplus \mathfrak{h}$. The involution implies the reductive structure $[\mathfrak{h},\mathfrak{h}]\subset\mathfrak{h}$, $[\mathfrak{h},\mathfrak{k}]\subset\mathfrak{k}$ and  $[\mathfrak{k},\mathfrak{k}]\subset\mathfrak{h}$. The dynamics are given by splitting the pullback of the Maurer-Cartan form $\grpel{V}^{-1}d\grpel{V}=\liel{P}+\liel{Q}\in\mathfrak{k}\oplus\mathfrak{h}$ and writing a lagrangian of the form $\Tr(\liel{P}\wedge\hodge \liel{P})$, which is invariant under the global symmetry group $G$ acting on the right of $\grpel{V}$.

We may choose the coset representative to satisfy a gauge, which geometrically amounts to specifying local sections of the bundle $G\stackrel{\pi}{\rightarrow}G/H$. We can keep the right action of $\grpel{}g\in G$ on $\grpel{V}$ in the gauge if we compensate it with a left action by an element $\grpel{}h\in H$ that depends on spacetime and $\grpel{V}$
\[\grpel{V}\longrightarrow \grpel{}{h}\,\grpel{V}\,\grpel{}{g}~.\]
Under this, $\liel{P}$ and $\liel{Q}$ transform as
\begin{align*}
 \liel{P}&\longrightarrow \grpel{}{h}\,\liel{P}\,\grpel{}{h}^{-1}\\
\liel{Q}&\longrightarrow \grpel{}h^{-1}d\grpel{}h+ \grpel{}{h}\,\liel{Q}\,\grpel{}{h}^{-1}~.
\end{align*}

The equations of motion are derived from an infinitesimal variation $\delta \grpel{V}=\epsilon(x) \grpel{V}$, where $\epsilon(x) \in\mathfrak{k}$ is arbitrary in any compact region. This gives
\[ d\left(\grpel{V}\hodge \liel{P}\,\grpel{V}^{-1}\right)=\grpel{V}\left(d\hodge \liel{P}-[\liel{Q},\hodge \liel{P}]\right)\grpel{V}^{-1}=0~,\]
or $\tilde{D}\hodge\liel{P}=d\hodge\liel{P}-[\liel{Q},\hodge \liel{P}]=0$. 
The Noether charges associated to $G$ are given by
\[\mathcal{C}=\int_{S^2}\grpel{V}\,\hodge \liel{P} \,\grpel{V}^{-1}\]
and, because of the equations of motion, they depend only on the homology of the surface $S^2$. If $\grpel{V}$ approaches at asymptotic infinity the trivial configuration, $\grpel{V}\rightarrow 1$, then the charges transform under $H$ in the representation of $\mathfrak{k}$. If we consider the transformations that preserve the asymptotic value of $\grpel{V}$, the action of the compensating $H$ on the charges is read from the asymptotic value $\grpel{}{h}\rightarrow \grpel{}{g}\in H$.

In our case the isotropy algebra is $\mathfrak{h}=\mathfrak{u}(1,1)$, which is the maximally non-compact subalgebra of $\su(1,2)$, see e.g.~\cite{Houart:2009ed}. This gives the non-riemannian symmetric space of signature $(++--)$ that we can see in the lagrangian of \eqref{eq:3dLag}. The Cartan complement $\mathfrak{k}$ transforms in the fundamental of $\mathfrak{u}(1,1)$. We will write quantities in the representation of $\mathfrak{u}(1,1)=\mathfrak{u}(1)\oplus \mathfrak{sl}(2,\RR)$ with generators $\liel{t}_a$ and $\liel{u}$
\begin{align*}
\liel{t}_x&\longmapsto \frac{1}{2}\begin{pmatrix}0&1\\1&0\end{pmatrix}\\
\liel{t}_y&\longmapsto \frac{1}{2}\begin{pmatrix}0&-i\\i&0\end{pmatrix}\\
\liel{t}_z&\longmapsto \frac{1}{2}\begin{pmatrix}-i&0\\0&i\end{pmatrix}\\
\liel{u} & \longmapsto \begin{pmatrix}i&0\\0&-i\end{pmatrix}
\end{align*}
and use a basis 
$\{\liel{e}_I=\liel{e}_v,\liel{e}_\omega, \liel{e}_\phi, \liel{e}_h\}$ 
of $\mathfrak{k}$ that is orthonormal with respect to the Killing form of $\su(1,2)$ and such that $a_1 \liel{e}_v+a_2 \liel{e}_\omega+b_1 \liel{e}_\phi+b_2 \liel{e}_h$ transforms under the representation of $\mathfrak{u}(1,1)$ as the vector
\[\begin{pmatrix}a_1+i\, a_2\\b_1+i\, b_2\end{pmatrix}~.\]

The lagrangian in \eqref{eq:3dLag} can be reproduced in the Borel gauge~\cite{Maison:1984tg}. The Isawawa decomposition $ \mathfrak{g}=\mathfrak{h}\oplus \mathfrak{b}$ defines the nilpotent Borel subalgebra $\mathfrak{b}=\mathfrak{a}\oplus\mathfrak{p}$, where $\mathfrak{a}$ is the maximal abelian algebra contained in $\mathfrak{k}$ and $\mathfrak{p}$ is the space spanned by the negative restricted roots of $\mathfrak{a}$. Here, $\mathfrak{a}$ is one-dimensional and spanned by the element $\liel{a}$, whereas $\mathfrak{p}$ contains a one-dimensional space spanned by an element $\liel{n}$ of root $-1$ and a two-dimensional space spanned by the elements $\liel{p}(1)$ and $\liel{p}(i)$ of root $-\frac{1}{2}$. In the appendix~\ref{app:su12} 
we write explicitly the representation of $\su(1,2)$ that we use. Then, choosing $\grpel{V}$ in the Borel subgroup as
\begin{align}
\grpel{V}&=\exp{(\log{V}\liel{a})}\exp{(\omega_0 \,\liel{n})}\exp{(\sqrt{2}\,\liel{p}(\phi+i h))}\nonumber\\\intertext{
gives the connection}
\label{eq:Qconn}
 \liel{Q}&=2\,V^{-1/2}\,d\phi\,\liel{t}_x+2 \, V^{-1/2}\,dh\, \liel{t}_y
+\frac{\omega_1}{V}\,\liel{t}_z-{\frac{3}{2}}\,\frac{\omega_1}{V}\, \liel{u}\\
\intertext{and the ``vielbein''}
 \liel{P}&=\frac{dV}{\sqrt{2}V}\,\liel{e}_v-\frac{\sqrt{2}\,\omega_1}{V}\,\liel{e}_w+\sqrt{2}V^{-1/2}\,d\phi\, \liel{e}_\phi+\sqrt{2}V^{-1/2}\, dh\, \liel{e}_h~.
 \label{eq:Pmom}
\end{align}
Plugging $\liel{P}$ into the lagrangian $\Tr\left(\liel{P}\wedge{\tilde\hodge}\,\liel{P}\right)$ matches the scalar sector of \eqref{eq:3dLag}. The charges $\mathcal{C}$ are matched to the Komar and field-strength definition as given in the equations \eqref{eq:Komardef1} and \eqref{eq:Komardef4}, and we find that they are represented by the vector
\begin{equation}\label{eq:Pcharges}
\mathcal{C}=\begin{pmatrix} M+iN\\-Q+iH\end{pmatrix} ~.
\end{equation}

Finally, the three-dimensional vielbein $\tilde{e}^i$ is inert under $SU(1,2)$.
 
\subsection{Reduction of Fermions}\label{sec:StatFerm}
In four dimensions the gravitino $\psi$ is a complex Rarita-Schwinger field and the supersymmetry parameters are complex spinor fields. Notice how electromagnetic duality is realized on the four-dimensional fermions. Say we use $F^*=\hodge F =-\Gamma_5\,F$ instead, so that $(g,F^*)$ is a new solution of the Einstein-Maxwell equations. The supersymmetry variation of the gravitino, equation \eqref{eq:deltapsi4}, is invariant if we transform $\eps\mapsto e^{-\frac{\pi}{4}\Gamma_5}\eps$ and $\psi\mapsto e^{-\frac{\pi}{4}\Gamma_5}\psi$. More generally a $U(1)$ rotation of $(F-i\,\hodge F)$ by $e^{i\theta}$ induces a rotation of the fermions by $e^{-\frac{\theta}{2}\Gamma_5}$.

In a sigma model, the bosonic sector is the metric $\tilde{g}$ on $\tilde{M}$ and the coset map  $\grpel{V}:\tilde{M}\rightarrow G/H$. The pullback of the bundle $G\stackrel{\pi}{\rightarrow} G/H$ by $\grpel{V}$ gives an  $H$-bundle over $\tilde{M}$ that we tensor with the spin bundle. The right action of $G$ on $\grpel{V}$ changes pointwise the fiber of the $\mathrm{Spin}\times H$-bundle via the compensating $H$ transformation. The fermions are sections of appropriate associated vector bundles of this latter bundle and transform under $G$ via the compensating $H$. The three-dimensional supersymmetry variations are covariant under the structure. We are interested to find the $H$-covariant fermions in three dimensions in terms of the timelike reduction of the four-dimensional fermions. We could have choosen to reduce the fermionic sector of the lagrangian for this, as was done for $\mathscr{N}=1$ supergravity in \cite{Gustafsson:1998ym}. However this is rather cumbersome and since the supersymmetry variations in three dimensions respect the coset structure, we will reduce the four-dimensional gravitino equation $\delta\psi=D\eps$. We can then identify the three-dimensional gravitino $\tilde{\psi}$, the gaugino $\tilde{\chi}$ and the three-dimensional supersymmetry parameter $\tilde\eps$ in terms of $\psi$ and $\eps$.

The Clifford algebra in three dimensions is isomorphic to the even part of the Clifford algebra in four dimensions, $\Cl(0,3)=\Cl\,{}^{\text{even}}(1,3)=\Mat(2,\CC)$, so that a Majoranna (or chiral) spinor reduces to a spinor of $\su(2)$. The complex spinors, a doublet of Majoranna spinors, reduce to a doublet of $\su(2)$ spinors and we will use the same symbol to describe them. We define the three-dimensional gamma matrices  $\gamma^{\underline{i}}=\Gamma^{\underline{0}\,\underline{i}}$ that satisfy $\{\gamma^{\underline{i}},\gamma^{\underline{j}}\}=-2\eta^{\underline{i}\,\underline{j}}$, where $\eta^{\underline{i}\,\underline{j}}=-\delta^{{\underline{i}\,\underline{j}}}$ is the flat metric. For convenience we also use the symbol $T=\Gamma^{\underline{0}}$. Note that in this representation of $\Cl(0,3)$ in $\Mat(4,\RR)$, the volume form $\gamma_5\equiv\Gamma_5=\gamma^{123}$ does not equal the imaginary unit. The imaginary unit still generates a rotation of the spinor doublet and commutes with the $\gamma^i$ and $T$, whereas $T$ and $\gamma_5$ provide the quaternionic structure of the $\su(2)$ spinors. We this notation we can keep the spinorial conventions we have presented so far without having to work with an explicit representation.

We assume that the timelike Killing vector $K$ leaves $\psi$ invariant. Likewise, we restrict to supersymmetry parameters $\eps$ that are invariant under $K$. The spinorial Lie derivative
\[\lie_K=\nabla_K-\frac{1}{4}dK^\flat\]
acts on a spinor field $\eps$ as an ordinary derivative $\partial_t$, which implies $\psi$ decomposes into the time-invariant $\psi_0$ and $\psi_i-B_i\psi_0$ (compare this with the decomposition of $A$ in \eqref{eq:stansatz}). However, the $H$-covariant fermions are a redefinition of these.

Let us explain our method. We correct the four-dimensional supersymmetry variations with an $\mathfrak{so}(1,3)$ transformation, so that they preserve the adapted form of the metric: $\iota_K e^{\underline{i}}=0$, $\underline{i}=1,2,3$. This is just the Lorentz boost $\Lambda_{\underline{i}\,\underline{0}}=-V^{-1/2}\rbs{\eps}{\Gamma_{\underline{i}}\,\psi_0}$. With this correction, we compute the variation of the three-dimensional vielbein, $\delta\tilde{e}\,{}^{\underline{i}}_j$, which is by assumption inert under $SU(1,2)$. We can bring $\delta\tilde{e}\,{}^{\underline{i}}_j$ to canonical form if we accompany it with an $\mathfrak{so}(3)$ rotation $\tilde{\Lambda}_{\underline{i}\,\underline{j}}=V^{-1/2}\rbs{\eps}{\Gamma_{{\underline{i}}\,\underline{j}\,\underline{0}}\,\psi_0}$. This allows us to identify the three-dimensional gravitino $\tilde{\psi}$ and supersymmetry parameter $\tilde{\eps}$. We then find how $\tilde{\psi}$ and $\tilde{\eps}$ transform under $U(1,1)$ by computing the variation $\delta\tilde{\psi}=\tilde{D}\tilde\eps$, reading the connection one-form of $\tilde{D}$ and comparing it with \eqref{eq:Qconn}. For this, we use the decomposition of the four-dimensional supercovariant derivative
\begin{align*}
D_0&=\lie_K-\frac{1}{4}dV+\frac{1}{2}\gamma_5\,\omega_1+\frac{i}{2}V^{1/2}T(d\phi-\gamma_5 dh)\\\intertext{and} D_i&=\tilde\nabla_i+\frac{1}{8V}[\gamma_i,dV]+\frac{V}{8}[dB,\gamma_i]-\frac{i}{2\sqrt{V}}T(d\phi-\gamma_5 dh)\gamma_i
\\&
+B_i\left(\frac{V^2}{4}dB-\frac{1}{4}dV+\frac{i}{2}V^{1/2}T(d\phi-\gamma_5 dh)\right) 
~,
\end{align*}
where we now use the three-dimensional Clifford algebra, e.g. $dV=\partial_i V\gamma^i$ is contracted with $\tilde{e}\,{}^{\underline{i}}_j$. The Lorentz corrections on the fermions are of order $O(\psi^2)$ and so we neglect them. 
Finally, it is easy to guess the gaugino $\tilde\chi$ in terms of $\psi_0$, so that its variation $\delta\tilde\chi$ is $U(1,1)$-covariant. 

A similar procedure was carried out in \cite{Gustafsson:1998ym} for the lagrangian reduction of $\mathscr{N}=1$ supergravity, to all orders in the fermions. We find the almost identical relations
\begin{align*}
\tilde\psi_i&=V^{1/4}\left(\psi_i-B_i\psi_0\right)+V^{-3/4}\gamma_i\,\psi_0\\
\tilde\chi&=V^{-3/4}\psi_0\\
\tilde\eps&=V^{1/4}\eps~.
\end{align*}
The supersymmetry variations of the vielbein and fermions are
\begin{align*}
\delta \tilde{e}\,{}^{\underline{i}}_j&=-\rbs{{\tilde{\eps}}}{T\gamma^{\underline{i}}\,\tilde{\psi}_j}\\
\delta \tilde\psi_i&=\tilde{D}_i(\liel{Q})\,\tilde{\eps}:=\left(\tilde\nabla_i-\frac{1}{2V}\gamma_5 (\omega_1)_i
+{i}V^{-1/2}\,T\, \partial_i\phi-{i}\,V^{-1/2}\,T\, \gamma_5\,\partial_i h\right)\tilde\eps\\
\delta\tilde\chi&=\gamma^i \liel{P}_i^I C_I\,\tilde\eps:=\left(-\frac{1}{4V}dV+\frac{1}{2V}\gamma_5\, \omega_1 +\frac{i}{2}V^{-1/2}T(d\phi-\gamma_5 \,dh)\right)\tilde\eps~.
\end{align*}
Comparing with the connection $\liel{Q}$ in \eqref{eq:Qconn}, we see that $\tilde\eps$ and $\tilde\psi$ transform under the compensating $\su(1,1)\subset\mathfrak{u}(1,1)$ in the representation $\hat{\liel{t}}_{x}=-\frac{i}{2}T$, $\hat{\liel{t}}_{y}=\frac{i}{2}T\gamma_5$ and $\hat{\liel{t}}_{z}=\frac{1}{2}\gamma_5$. They do not transform under the $\mathfrak{u}(1)\subset \mathfrak{u}(1,1)$. We have also defined the $C_I$ matrices
\begin{align*}
 C_v&=-\frac{\sqrt{2}}{4}&C_w&=-\frac{\sqrt{2}}{4}\gamma_5\\
 C_\phi&=-\frac{i\sqrt{2}}{4}T&C_h&=\frac{i\sqrt{2}}{4}T\gamma_5~.
\end{align*}
They satisfy $f_{aI}{}^J C_J=-C_I \hat{\liel{t}}_a$, where $f_{aI}{}^J$ are the structure coefficients of the $\mathfrak{su}(1,1)$ action on $\mathfrak{k}$: $[\liel{t}_a,\liel{e}_I]=f_{aI}{}^J \liel{e}_J$. Therefore, the gaugino is invariant under the compensating $\su(1,1)$. It transforms however under the compensating element $\liel{u}\in \mathfrak{u}(1)$ as $\delta\tilde\chi=-\gamma_5\tilde\chi$.

\subsection{Stationary BPS solutions}\label{sec:StatBPS}
The Reissner-Nordstr\"om black holes are characterized by their mass $M$, electric charge $Q$ and magnetic charge $H$. They can be generalized to a family of solutions characterized also by a non-zero topological NUT charge $N$~\cite{PhysRev.133.B845}. The integrability condition of a Killing spinor requires that their charges saturate the BPS bound $M^2+N^2=Q^2+H^2$. The local form of the metric and field-strength for the $1/2$-BPS solutions is
\begin{subequations}\label{eq:GenNUT}
\begin{align}
g&=\frac{(r-M)^2}{R^2}(dt+2N\cos\theta \,d\phi)^2-\frac{R^2}{(r-M)^2}\,dr^2-R^2\,ds^2(S^2)\\
F&=\frac{Q(r^2-N^2)-2\,r\,N\,H}{R^4}\,dr\wedge dt+\frac{H(r^2-N^2)+2\,r\,N\,Q}{R^2}\,\dvol_{S^2}~,
\end{align}\end{subequations}
where $R^2=r^2+N^2$ and $ds^2(S^2)$ is the metric element of a unit two-sphere. Let us use the notation of~\cite{Argurio:2008zt} and define the angles $a_m$, $a_q$ and $\beta(r)$ via the relations 
\begin{align*}
R\,e^{i\,\beta}&=r+i \,N~,&\mathcal{Z}\,e^{i \, a_m}&=M+i\, N ~, & \mathcal{Z}\,e^{i \, a_q}&=-Q+i\, H~, 
\end{align*}
where $\mathcal{Z}=\sqrt{M^2+N^2}$. It then becomes convenient to write the field-strength in the Clifford algebra\footnote{the Clifford algebra $\Cl(1,3)=\Mat(4,\RR)$ is isomorphic as a vector space to the exterior algebra and has a unique faithful irreducible representation that we use.} as
\begin{equation}\label{eq:NUTFinCl}
F=\mathcal{Z}\,R^{-2}\, \Gamma^{\underline{0}\,\underline{r}}\,\exp{\left((a_q-2\beta)\Gamma_5\right)}~.
\end{equation}
In the appendix~\ref{app:D} we expand the supercovariant derivative $D$ in our conventions. The solution to the Killing spinor equation $D_\mu\eps=0$ is given by
\[\eps=
e^{-\frac{1}{2}\Gamma^{\underline{\theta}\, \underline{r}}\theta}
e^{\frac{1}{2}\Gamma^{\underline{\theta}\,\underline{\phi}}\phi}
\left(\frac{r-M}{R}\right)^{\frac{1}{2}}
e^{-\frac{\beta}{2}\Gamma_5}\eps_0~,\]
where $\eps_0$ is a constant spinor that satisfies the BPS projection
\begin{equation}\label{eq:BPSproj}
\left(1+i\, e^{(a_q-a_m)\Gamma_5}\Gamma_{\underline{0}}\right)\eps_0=0~.
\end{equation}
A basis for the Killing spinors can be labelled, e.g., by the real part of their asymptotic value $\{\eps_{\mathcal{I}}\}_{\mathcal{I}=1,2,\ldots 4}$.

A Killing spinor $\eps$ of the background squares to a Killing vector $\rbs{\eps}{\Gamma^\mu\eps}\partial_\mu$ that, due to the BPS projection, is proportional to the timelike isometry $K=\partial_t$. In turn, $K$ acts trivially on the Killing spinors, $\lie_K\eps=0$. The metric and field-strength are also preserved by three spacelike Killing vectors
\begin{align*}
\xi_x&=\sin\phi\frac{\partial}{\partial\theta}+\cos\phi\cot\theta\frac{\partial}{\partial\phi}-2N\frac{\cos\phi}{\sin\theta}\frac{\partial}{\partial t}\\
\xi_y&=\cos\phi\frac{\partial}{\partial\theta}-\sin\phi\cot\theta\frac{\partial}{\partial\phi}+2N\frac{\sin\phi}{\sin\theta}\frac{\partial}{\partial t}\\
\xi_z&= \frac{\partial}{\partial\phi}
\end{align*}
that form a representation of $\su(2)$ and commute with $K$. The spinorial Lie derivative along $\xi_x$ and $\xi_z$ is
 found to be
\begin{align*}
\lie_{\xi_x}&=\sin\phi\,\partial_\theta+\cos\phi\cot\theta\,\partial_\phi-2N\frac{\cos\phi}{\sin\theta}\partial_t+\frac{1}{2}\frac{\cos\phi}{\sin\theta}\Gamma^{\underline{\phi}\,\underline{\theta}}\\
\lie_{\xi_z}&=\frac{\partial}{\partial\phi}~.
\end{align*}
From this and using the Baker-Campbell-Hausdorff formula, we find the action of the $\su(2)$ on the Killing spinors\footnote{only their $\theta$, $\phi$ and $t$ dependence is relevant here.}. It induces an action of $\xi_x$, $\xi_y$ and $\xi_z$ on $\eps_0$ and hence on the basis $\{\eps_{\mathcal{I}}\}$, which is given by  $\frac{1}{2}\Gamma^{\underline{\theta}\,\underline{\phi}}$, $\frac{1}{2}\Gamma^{\underline{r}\,\underline{\theta}}$ and $\frac{1}{2}\Gamma^{\underline{\phi}\,\underline{r}}$, respectively.

The above discussion describes the structure of the Killing superalgebra $\su(2)\oplus \mathrm{span}\left\langle K\right\rangle \oplus\mathrm{span}\langle\eps_{\mathcal{I}}\rangle$ of a NUT-charged BPS black hole. If we choose the symplectic inner product of Majoranna spinors to be $\mathit{C}=\Gamma^{\underline{0}}$, the Killing superalgebra coefficients in our chosen basis are given by
\begin{align*}
 [\eps_{\mathcal{I}},\eps_{\mathcal{I}'}]&=-2\,\delta_{\mathcal{I}\mathcal{I}'}K\\
[K,\eps_{\mathcal{I}}]&=0&[K,\xi_i]&=0&
[\xi_i,\xi_j]&=\epsilon_{ijk}\,\xi_k\\
[\xi_x,\eps_{\mathcal{I}}]&=\frac{1}{2}\left(\Gamma^{\underline{\theta}\,\underline{\phi}}\right)_{\mathcal{I}}{}^{\mathcal{J}}\eps_{\mathcal{J}}&
[\xi_y,\eps_{\mathcal{I}}]&=\frac{1}{2}\left(\Gamma^{\underline{r}\,\underline{\theta}}\right)_{\mathcal{I}}{}^{\mathcal{J}}\eps_{\mathcal{J}}&
[\xi_z,\eps_{\mathcal{I}}]&=\frac{1}{2}\left(\Gamma^{\underline{\phi}\,\underline{r}}\right)_{\mathcal{I}}{}^{\mathcal{J}}\eps_{\mathcal{J}}~.
\end{align*}
We observe that BPS solutions of distinct charges have isomorphic Killing superalgebras. That is, we can always give the Killing brackets in the basis above. By extending the Killing superalgebra using the coset structure in three dimensions and embedding it into  flat space, we will be able to give a $U(1,1)$-covariant superalgebra that distinguishes the various configurations. Of course, the Killing superalgebra extensions of different charged configurations will still be isomorphic, the isomorphism been given by the $SU(1,1)$ compensating transformation.

As observed by Misner~\cite{Misner:1963fr}, the NUT-charged solutions do not have a \emph{globally} flat limit. One way to see this is to note that, for $N\neq 0$, the poly-vector $\xi_x\wedge\xi_y\wedge\xi_z$ does not vanish and the action of $SU(2)$ traces a three-sphere at any constant $r$. Evidently, the manifold has closed timelike curves and there are no spacelike hypersurfaces, for the latter see the argument in~\cite{Misner:1963fr}. At best, one introduces a Dirac string and joins two hemispheres, with local metric and field-strength given by \eqref{eq:GenNUT}, under the diffeomorphism $t_\pm\mapsto t\pm N d\phi$. This makes it unpleasant to compare a BPS solution with flat space in four dimensions, e.g. by looking at the region $r\rightarrow \infty$. One would like to interpret the Killing spinors as a subspace of Minkowski's space Killing spinors in a certain smooth limit. But here the limit $r\rightarrow\infty$, or equivalently $(M,N,Q,H)\rightarrow 0$, changes the topology of the space\footnote{note that the diffeomorphism used to patch the two hemispheres is a time translation, whereas the Killing spinors are time independent. This means that the Killing spinors are globally defined. It also means that they can be reduced to spinors on the three-dimensional space of orbits.}. Furthermore, an ADM definition of mass requires a spacelike hypersurface that is lacking here\footnote{alternatively, one cuts out the tubes around the Dirac strings, see e.g.~\cite{Hunter:1998qe}. But the reference background should still have the same topology.}. 

Nevertheless, we will make sense of the Nester-Witten form, which is otherwise used to express the ADM mass. We show that is agrees with Komar's definition of mass that is well-defined for $N\neq 0$. We will see that the Nester-Witten form has a natural generalization in three-dimensions that is covariant under the coset structure. The embedding of the NUT-charged black holes in flat space is straightforward in three dimensions, where the reduced metric $\tilde{g}$ is asymptotically (globally) flat and the coset $\grpel{V}$ approaches the trivial configuration. For this reason, the extension of the Killing superalgebra and its embedding in an extension of the Poincar\'e superalgebra is most relevant in three dimensions.

In the reduced theory, the $1/2$-BPS family of solutions lie on the same orbit of the right $U(1,1)\subset SU(1,2)$ action. We can easily translate quantities into three-dimensional ones. The BPS projection is automatically written in our conventions, by replacing $\Gamma_5$ and $\Gamma_{\underline{0}}$ with $\gamma_5$ and $T$, respectively. We can use the Clifford algebra to compute $d\phi=\iota_K F = -\frac{1}{2}V^{1/2}[T,F]$ and $dh=\iota_K\hodge F=\frac{1}{2}V^{1/2}[T,\gamma_5 F]$. By using relations of the form
\[(r-iN)(M-r+r+iN)=\mathcal{Z}\,R\exp((-\beta+a_m)i)~,\]
one can write the ``vielbein'' $\liel{P}$ in the form
\begin{multline}
 \liel{P} =\sqrt{2\,V} \mathcal{Z} \left(\cos(-\beta+a_m)\,\liel{e}_v+\sin(-\beta+a_m)\,\liel{e}_w\right.\\\left.+\cos(-2\beta+a_q)\,\liel{e}_\phi+\sin(-2\beta+a_q)\,\liel{e}_h\right){\tilde\hodge}\,\dvol(S^2)~.
\end{multline}
The charges are then represented by the vector
\[\mathcal{C}=\begin{pmatrix} M+iN\\-Q+iH\end{pmatrix}~,\]
as expected. 
Note how the action of $\mathfrak{u}(1,1)$ preserves the BPS condition $M^2+N^2=Q^2+H^2$ but rotates the angles $a_m$ and $a_q$ and scales $\mathcal{Z}=\sqrt{M^2+N^2}$. The element $\liel{u}\in\mathfrak{u}(1)\subset\mathfrak{u}(1,1)$ rotates the two angles equally, $\delta a_q=\delta a_m$. The Killing spinors do not transform under $\liel{u}$ and this is reflected in the BPS condition, equation \eqref{eq:BPSproj}, which depends only on the phase $a_m-a_q$. The BPS saturated charges are fixed by an element $\liel{Z}_0$ in $\su(1,1)$. This element belongs to the extended Killing superalgebra, as we show in section~\ref{sec:SusyNW}. Finally, the three-dimensional metric is easily seen to be globally flat for $r>M$.

\section{Supersymmetry Algebra}\label{sec:Susy}
In supergravity theories the commutator of two supersymmetry transformations closes on the even symmetries {of the theory}. These are diffeomorphisms, gauge shifts and local Lorentz transformations. This is the field-theoretic origin of the Killing superalgebra of a background, $\mathfrak{l}=\mathfrak{l}_0\oplus\mathfrak{l}_1$. A Killing spinor $\eps\in \mathfrak{l}_1$ squares to a vector field, say  $\xi=\rbs{\eps}{\Gamma^\mu\eps}\partial_\mu\in\mathfrak{l}_0$, which leaves the metric and field-strength invariant, $\lie_\xi g=\lie_\xi F=0$. The even-odd bracket of the Killing superalgebra is defined via the spinorial Lie derivative. It takes Killing spinors into Killing spinors, because the Lie derivative along a Killing vector preserves the kernel of $D:\eps\mapsto D\eps$. The even-even bracket is defined by the Lie bracket of vector fields, which closes on Killing vectors. The Killing superalgebra of a background is the restriction of the superalgebra of the theory to the background's Killing spinors and Killing vectors. Note that the supersymmetry transformations of the theory use anticommuting spinor fields. Relating this to the Killing superalgebra of a background requires that we mod the Grassmann-odd parity of the spinors, so that the odd-odd bracket of the latter is symmetric.

We can extend the symmetries of a background by taking into account some of the gauge-dependent fields. Already in $\mathcal{N}=2$ we can use the symmetries of the gauge field $A$ and its dual $A^*$. If a vector $\xi$ preserves $F$, it preserves $A$ up to a gauge shift $\Xi_A$
\[\lie_\xi A= -d\left(\Xi_A-\iota_\xi A\right)~.\]
We shifted the scalar $\Xi_A$ by $-\iota_\xi A$ so that $\Xi_A$ is globally defined. That is, in a patch $U_{\alpha\beta}$ where the gauge field is described by $A_\alpha$ and $A_\beta$, with $A_\alpha-A_\beta=d\chi_{\alpha\beta}$, the scalar $\Xi_A$ is gauge invariant. Note that for non-trivial configurations, the scalar $\Xi_A$ is generically non-zero. This can be checked, e.g., for the Dirac monopole in flat space. The commutator of two transformations 
\[\delta^2_i A = 
\lie_{\xi_i} A + d\left(\Xi_A{}_i-\iota_{\xi_i} A\right)~,\quad i=1,2~,\]
that preserve $A$, gives 
\[
[\delta^2_1,\delta^2_2]A= \lie_{[\xi_1,\xi_2]} A + d\left(\lie_{\xi_1}\Xi_A{}_2-\iota_{[\xi_1,\xi_2]} A\right)~.\]
The Lie bracket on $(\xi,\Xi_A)$ that is implied can be shown to be antisymmetric and satisfy the Jacobi identity.

The description of the bosonic symmetries alone by using doublets of the form $(\xi,\Xi_A)$ is redundant for the following reason. The symmetry algebra of a background is a finite-dimensional space. We can thus label the doublets by the Killing vector $\xi$. The difference of two such doublets that share the same Killing vector $\xi$ is proportional to a trivial gauge shift, whose addition into the Lie algebra is inconsequential. However, the inclusion of the odd symmetries restores the presence of this element in the odd-odd bracket. The square of a supersymmetry transformation acts on $A$ as
\[\delta^2 A=\lie_\xi A +d\left(-\rbs{\eps}{i\,\eps}-\iota_\xi A\right)~,\]
where $\xi=\rbs{\eps}{\Gamma^\mu\eps}\partial_\mu$. 
Similarly, it acts on the dual gauge field $A^*$, with ${\hodge} F=dA^*$, as
\[\delta^2 A^*=\lie_\xi A^* +d\left(\rbs{\eps}{i\,\Gamma_5 \eps}-\iota_\xi A\right)~.\]

Let us tentatively\footnote{we have not written the even-odd or even-even brackets for a generic background. We will do so for stationary backgrounds and show it is consistent.}
 call this superalgebra, whose odd-odd bracket on a symmetric bispinor field $\eps\times\eps$ gives us the vector field $\xi$ and the scalar fields $\Xi_A$ and $\Xi_{A^*}$, the electromagnetic superalgebra of the theory. By restricting it to the symmetries of a background, we shall call it the electromagnetic Killing superalgebra of the background. 
The odd-odd bracket of the electromagnetic Killing superalgebra of flat space, whose Killing spinors are the constant Dirac spinors, is surjective on $\RR^{1,3}\oplus\RR\oplus\RR$. This is reflected, in a more familiar form, in the superalgebra of charges for an asymptotically flat space
\begin{equation}\label{eq:EMPoinc}
[Q^i_\alpha,Q^j_\beta]=
\delta^{ij}\left(\Gamma^\mu C^{-1}\right)_{\alpha\beta} {P}_\mu 
+ \epsilon^{ij}C_{\alpha\beta}\, {Q}
+\epsilon^{ij}\left(\Gamma_5\,C\right)_{\alpha\beta} {H}~.
\end{equation}

The electromagnetic Killing superalgebra is covariant under electromagnetic duality $F\mapsto e^{\theta\Gamma_5}F$,  whereby the Killing spinors transform as $\eps\mapsto e^{\frac{\theta}{2}\Gamma_5}\eps$ and $(\Xi_{A}+i\,\Xi_{A^*})\mapsto e^{i\theta}(\Xi_{A}+i\,\Xi_{A^*})$. Furthermore, the electromagnetic superalgebra of charges in \eqref{eq:EMPoinc} is invariant under the transformation with $(Q+i\,H)\mapsto e^{-i\theta}(Q+i\,H)$. Given an isometry $K$, the symmetry of the theory is enlarged to $SU(1,2)$. All the information of the various charged configurations can still be extracted from the same electromagnetic superalgebra. As an example take the $1/2$-BPS solution, whose Killing spinors satisfy the BPS projection of \eqref{eq:BPSproj}. Then the right-hand side of \eqref{eq:EMPoinc} restricted to the Killing spinors and the corresponding BPS charges vanishes. However, the electromagnetic Killing superalgebra and the superalgebra of charges are not covariant under $SU(1,2)$. In other words, the charge $N$ does not appear in \eqref{eq:EMPoinc} on an equal footing to the charges $M$, $Q$ and $H$. We will remedy this in section \ref{sec:SusyNW}.

\subsection{The SU(1,1) triplet}\label{sec:SusyTriplet}
A stationary background defines the connection $\theta=dt+B$ on the fiber bundle defined by the continuous isometry. The connection splits any vector field $\xi$ into its direction along the fiber $\theta(\xi)\,K$ and its complement $\tilde{\xi}=\xi-\theta(\xi)\,K$. If $\xi$  commutes with $K$, then $\theta(\xi)$ and $\tilde\xi$ are well-defined on the base space. Note that, although $\tilde{\xi}$ has a component in the direction of $\partial_t$, the ``gauge'' condition $\theta(\tilde\xi)=0$ fixes it completely. We will use the same symbol for the vector $\tilde{\xi}$ on the base space.

A natural question to ask is how does supersymmetry act on $B$. In section~\ref{sec:StatFerm} we corrected the supersymmetry transformation under a spinor field $\eps$ with a local Lorentz transformation, in order to preserve the gauge $\iota_K e^{\underline{i}}=0$. The square of the supersymmetry transformation acts on the vielbeins $e^{\underline{\mu}}$ via an infinitesimal diffeomorphism $\lie_\xi$ and a local Lorentz transformation, whose boost component is fixed to be $\Lambda^{\underline{i}}{}_{\underline{0}}=-\frac{1}{\sqrt{V}}\iota_K \lie_\xi e^{\underline{i}}$. For a time-independent spinor this is zero, \[\iota_K\lie_\xi e^{\underline{i}}=\lie_\xi \iota_K e^{\underline{i}}-\iota_{[\xi,K]}e^{\underline{i}}=0~.\]
The square of the supersymmetry transformation on $\theta=\frac{1}{\sqrt{V}}e^{\underline{0}}$ is thus given by $\lie_\xi\theta$. From this, we extract the square of a supersymmetry transformation of $B$
\[\delta^2 B = \lie_{{\xi}}B+d\left(\Xi_B-\iota_\xi B\right)~,\]
where \[\Xi_B=\theta(\xi)=\rbs{\eps}{\Gamma^{\underline{0}}\,\eps}~.\]
Just as supersymmetry incorporates in the odd-odd bracket the scalars $\Xi_A$ and $\Xi_{A^*}$, whose importance we noticed is at most a trivial gauge transformation, it incorporates the scalar $\Xi_B$ that is related to a trivial translation along $K$.

We motivated the scalars $\Xi_A$, $\Xi_{A^*}$ and $\Xi_{B}$ as gauge shifts of respectively $A$, $A^*$ and $B$. These gauge fields are then dualized to scalars that, along with $V$, parametrize the coset $SU(1,2)/U(1,1)$. The triplet $(\Xi_A,\Xi_{A^*},\Xi_{B})$ does not play a role in the symmetries of the coset scalars. Indeed, in three-dimensions the square of supersymmetry transformations on the coset scalars closes on three-dimensional diffeomorphisms alone. The triplet nevertheless does appear in the symmetric tensor square of the three-dimensional spinor fields
\begin{equation*}
 (Z_x,Z_y,Z_z)=\left(\rbs{\tilde\eps}{i\,\gamma_5\,\tilde\eps},\rbs{\tilde\eps}{i\,\tilde\eps},\rbs{\tilde\eps}{T\,\tilde\eps}\right)~.
\end{equation*}
This triplet transforms in the co-adjoint of the compensating transformation of $SU(1,1)$. The Killing form of $\su(1,1)$ has signature $(-,+,+)$ on $(\liel{t}_z,\liel{t}_x,\liel{t}_y)$. The $\su(1,1)$ element 
\begin{equation}\label{eq:defZ}
 \liel{Z}=Z^i\,\liel{t}_i=-\rbs{\tilde\eps}{i\,\gamma_5\,\tilde\eps}t_x+\rbs{\tilde\eps}{i\,\tilde\eps}t_y+\rbs{\tilde\eps}{T\,\tilde\eps}t_z
\end{equation}
 transforms, under $\grpel{V}\mapsto h\,\grpel{V}\,g$, as $\liel{Z}\mapsto h\,\liel{Z}\,h^{-1}$. It remains inert under the $U(1)$.

Any Killing spinor bilinear naturally obeys a differential property. We can easily check that the $\su(1,1)$ triplet $\liel{Z}$ generated from Killing spinors satisfies the gauge covariant equation \[\tilde{D}\liel{Z}=d\,\liel{Z}-[\liel{Q},\liel{Z}]=0~.\]
Interestingly, we also observe that the four-dimensional Fierz identity
\[ g(\xi,\xi)=\rbs{\eps}{i\,\eps}^2+\rbs{\eps}{i\,\Gamma_5\,\eps}^2\,\]
which holds for any Dirac spinor field $\eps$, 
becomes a statement about the $\mathfrak{su}(1,1)$ invariant
\[ \tilde{g}(\tilde{\xi},\tilde{\xi})=-Z_z^2+Z_x^2+Z_y^2~,\]
where the vector $\tilde\xi=\rbs{\tilde\eps}{T\gamma^i\tilde{\eps}}\partial_i$ is inert under $\mathfrak{su}(1,1)$.

The supersymmetry parameters of the reduced theory transform under the compensating $SU(1,1)$. The symmetric square of such a spinor field gives a $U(1,1)$-invariant vector field and an $\su(1,1)$ scalar. The restriction to Killing spinors closes on three-dimensional Killing vectors and gauge-covariant constant $\su(1,1)$ scalars. We define the odd-odd Killing bracket 
\begin{equation}\label{eq:3dooKbrack}
[\tilde\eps,\tilde\eps]=\tilde\xi+\liel{Z}~,\end{equation}
where $\tilde\xi=\rbs{\tilde\eps}{T\gamma^i\tilde{\eps}}\partial_i$ is the three-dimensional Killing vector and $\liel{Z}$ is the gauge-covariant constant scalar of $\su(1,1)$ as in \eqref{eq:defZ}. The odd-odd Killing bracket in \eqref{eq:3dooKbrack} transforms equivariantly under the compensating $SU(1,1)$ and is invariant under the compensating $U(1)$.

\subsection{Killing brackets}\label{sec:SusyBrack}
We will here postulate suitable even-odd and even-even brackets. Given a Killing vector $\tilde\xi$ and a gauge-covariant constant $\liel{Z}$, the two natural operations on a Killing spinor $\tilde\eps$ 
\begin{align*}
 \tilde\eps & \mapsto \tilde{\lie}_{\tilde{\xi}} \tilde\eps = (\tilde\nabla_{\tilde{\xi}}-\frac{1}{4}(d\tilde\xi)_{ij}\gamma^{ij})\,\tilde\eps\\\intertext{and} \tilde\eps &\mapsto \liel{Z}\,\tilde\eps=Z^i \,\hat{\liel{t}}_i \, \tilde\eps
\end{align*}
both 
give a new Killing spinor. However, the second choice will not in general satisfy the odd-odd-odd Jacobi identity. As a counterexample, the odd-odd-odd Jacobi identity for the constant spinors of flat space would require that the expression
\[ \left(-\overline{\eps}_1{\Gamma^{\underline{0}}\,\eps_1}\Gamma_5  -\overline{\eps}_2{\Gamma^{\underline{0}}\,\eps_2}\Gamma_5
-2\overline{\eps}_1{\eps_2}i\,\Gamma^{\underline{0}}\,\Gamma_5+2\overline{\eps}_1{\Gamma_5\,\eps_2}\,i\,\Gamma_5\right)\left(\eps_1+i\,\eps_2\right) \]
vanishes for arbitrary Majoranna spinors $\eps_1$ and $\eps_2$. This cannot hold because the expression is inhomogeneous in $\eps_1$ and $\eps_2$. The first choice, the action of the Lie derivative, is the same action induced by the four-dimensional $\eps\mapsto \lie_\xi\eps$, with  $\xi=\tilde{\xi}+\theta(\xi)\,K$ and $\eps=V^{-1/4}\tilde{\eps}$. Note that the manifestly gauge covariant  $\tilde\eps\mapsto(\tilde{\lie}_{\tilde{\xi}}-\liel{Q}_{\tilde{\xi}})\,\tilde\eps$  agrees with the Lie derivative action because  $\liel{Q}_{\tilde{\xi}}=0$.

In fact all Killing vectors that are generated from Killing spinors act trivially on the Killing spinors. Indeed, for $\tilde\xi$ generated by $\tilde\eps$, we have
\begin{equation*}
\tilde\nabla_i\tilde\xi_j=2\,\rbs{\tilde\eps}{T\,\gamma_j\,\tilde\nabla_i\tilde\eps}
=2\,\,\rbs{\tilde\eps}{T\,\gamma_j\,(-Q^x_i\,\frac{i}{2}\,T + Q^y_i \frac{i}{2}T\,\gamma_5 + Q^z_i\, \frac{1}{2} \,\gamma_5)\tilde\eps}=0~,\end{equation*}
which vanishes because of antisymmetry in the spinor bilinear. Then, with $\tilde{\eps}'$ another Killing spinor, $\tilde{\lie}_{\tilde\xi}\tilde\eps'=\tilde\nabla_{\tilde\xi}\tilde\eps'=\tilde{D}_{\tilde\xi}\tilde\eps'=0$ proves our assertion. The same is not true with Killing vectors that are not in the image of the odd-odd bracket. For instance, the Killing spinors of the BPS solutions transform in a representation of $\su(2)$ under the $\tilde\xi_i$. 

The action of a Killing vector $\tilde\xi$ on $\liel{Z}$ is implied by the even-odd-odd Jacobi identity, $\liel{Z}\mapsto \tilde\nabla_{\tilde\xi}\liel{Z}$, but this is again zero since $\tilde\nabla_{\tilde\xi}\liel{Z}=\tilde{D}_{\tilde\xi}\liel{Z}=0$. We conclude that the Killing spinors generate only central elements. Therefore, all the information of the Killing superalgebra ideal $\mathfrak{l}_1\oplus S^2\mathfrak{l}_1$, where $\mathfrak{l}_1$ is the space of Killing spinors, is in the odd-odd bracket, equation \eqref{eq:3dooKbrack}. 

The odd-odd bracket of the Killing superalgebra of the BPS backgrounds of section \ref{sec:StatBPS} is extended to
\[ [\eps_{\mathcal{I}},\eps_{\mathcal{I}'}]=-2\delta_{\mathcal{I}\mathcal{I}'}\,\liel{Z}~,\]
where
\[\liel{Z}=\liel{t}_z+\sin(a_q-a_m)\,\liel{t}_x-\cos(a_q-a_m)\,\liel{t}_y~.\]
We have not shown how $SU(1,1)$ acts on the basis $\eps_{\mathcal{I}}$ but, by our natural construction, the odd-odd bracket transforms covariantly under $SU(1,1)$. Before we move on to discuss the four-dimensional asymptotically \emph{locally} flat backgrounds, let us make clear that the results obtained so far are generic for stationary backgrounds. If the background is asymptotically \emph{globally} flat in the reduced theory, then the extended Killing superalgebra that we have discussed is a subalgebra of the one of flat spacetime's. This is true as it is a matter of taking a well defined limit of the background in three dimensions. In the next section we discuss how this ties in with the extended superalgebra of charges, the Nester-Witten formula and the NUT charge.

\subsection{Global Superalgebra}\label{sec:SusyNW}
Asymptotically flat backgrounds in four dimensions have a well-defined notion of the total four-momentum $P^{\underline{\mu}}$ of the space~\cite{Arnowitt:1962hi}, which transforms covariantly under the asymptotic symmetry group $SO(1,3)$. A simple expression for $P^{\underline{\mu}}$ using Dirac spinors was given initially by Witten  and shortly later refined by Nester~\cite{Witten:1981mf,*Nester:1982tr}. If a Dirac spinor field $\eps$ approaches the constant value $\eps_0$ at the asymptotic boundary of spatial infinity $S^2_\infty$, then
\[P^{\underline{\mu}}\,\rbs{\eps_0}{\Gamma_{\underline{\mu}}\,\eps_0} = \int_{S^2_\infty} \overline{\eps}\,{\Gamma_5\Gamma_\mu \nabla_\nu\eps}\, dx^\mu\wedge dx^\nu + \text{c.c.}~.\]
The existence of Dirac-Witten spinors, which satisfy $\sum_{i=1}^3\Gamma^i D_i\eps=0$, implies the positivity of the ADM mass $P^{\underline{\mu}} P_{\underline{\mu}}\geq 0$. A more appropriate form for Einstein-Maxwell's theory is given by Gibbons and Hull~\cite{Gibbons:1982fy}:
\[L_0^\CC=\overline{\eps}\,{\Gamma_5\Gamma_\mu D_\nu\eps} \, dx^\mu\wedge dx^\nu~.\]
Expanding the connection $D_\mu=\nabla_\mu-\frac{i}2 F\,\Gamma_\mu$ and integrating the real part of $L_0^\CC$ over $S^2_\infty$ gives
\begin{equation*}
\int_{S^2_\infty} \text{Re}(L^\CC_0) =
\rbs{\eps_0}{\Gamma_{\underline{\mu}}\,\eps_0}P^{\underline{\mu}} - \rbs{\eps_0}{ \mathrm{i}\,\Gamma_5\,\eps_0} H -\rbs{\eps_0}{ \mathrm{i}\,\eps_0}Q~.
\end{equation*}
This can be used to show a stricter bound, $M^2\geq Q^2+H^2$, using a generalized Dirac-Witten condition. These formulae are valid provided the spacelike $S^2$ and the limiting $S^2_\infty$ exist.

There is a nice connection between the Nester-Witten formula and the asymptotic superalgebra of charges~\cite{Hull:1983ap}. The total supercharge of a background
\[ \rbs{\eps_0}{Q[e^{\underline{\mu}},\psi]}=\int_{S^2_\infty} \overline{\eps}\,{\Gamma_5 \Gamma_\mu \psi_\nu} \, dx^\mu\wedge dx^\nu +\text{c.c.}~\]
is formally a functional of the vielbein and gravitino that generates supersymmetry transformations via the supergravity Dirac brackets~\cite{Teitelboim:1977hc}. Recall that the real spinor inner product is symplectic on Dirac spinors. We can therefore decompose the supercharge into the Majoranna basis $Q^{\mathcal{I}}_\alpha$, $\mathcal{I}=1,2$ and $\alpha=1,2,3,4$. The odd-odd bracket of supercharges can be computed by transforming the supercharge under a supersymmetry transformation. This gives the odd-odd bracket
\begin{equation}\label{eq:EMPoinc2}
 [Q^{\mathcal{I}}_\alpha,Q^{\mathcal{J}}_\beta]= \delta^{{\mathcal{I}}{\mathcal{J}}}\left(\Gamma^{\underline{\mu}} C^{-1}\right)_{\alpha\beta} {P}_{\underline{\mu}} + \epsilon^{{\mathcal{I}}{\mathcal{J}}}(C^{-1})_{\alpha\beta} {Q}  +\epsilon^{{\mathcal{I}}{\mathcal{J}}}\left(\Gamma_5\,C^{-1}\right)_{\alpha\beta} {H}~.
\end{equation}
Now, interpret a BPS charged black hole as a stable quantum state above the flat space vacuum. The BPS state preserves only a fraction of the vacuum's supersymmetry, which is the space of Killing spinors, as a subspace of the constant spinors at asymptotic infinity. Indeed, the right-hand side of the extended Poincar\'e superalgebra of charges restricted to the BPS Killing spinors, vanishes when we let it act on the BPS state with the eigenvalues $\hat{P}_\mu=(M,0,0,0)$, $\hat{Q}=Q$ and $\hat{H}=H$.

The vanishing of the left-hand side of \eqref{eq:EMPoinc2} holds invariably for the NUT-charged black holes of section~\ref{sec:StatBPS} as well, by taking the limit $r\rightarrow\infty$ and using the BPS projection of the Killing spinors. These backgrounds however are not asymptotically flat, neither do they have a spacelike hypersurface. This makes it problematic to define the global charges, e.g. the mass and supercharge, by using an ADM $3+1$ decomposition. Furthermore, the proof of the Bogomolny inequality by Gibbons and Hull requires the existence of Dirac-Witten spinors on a spacelike hypersurface. This is why the Komar mass $M$ can be made arbitrary small by compensating the value of $N$. The Bogomolny inequality is corrected to $M^2+N^2\geq Q^2+H^2$, see~\cite{Heusler:1997am}, which the BPS states saturate. That being said, on these {stationary} backgrounds it is trivial to construct time-independent global quantities. More importantly, these quanitites should be a suitable generalization of the zero NUT-charged backgrounds that are covariant under the coset structure. It is these quantities that we call covariantized under the coset structure. 

Let us momentarily drop the use of the symplectic form in favour of the antihermitian product, the two being related by
\[ \overline{\eps}{\eps'}=\frac{1}{2}\left(\rbs{\eps}{\eps'}-\mathrm{i}\,\rbs{\eps}{\mathrm{i}\,\eps'}\right)~.\]
We begin by reducing to three dimensions the complex two-form 
\[L^\CC_1=\overline{\eps}\,{\Gamma_5 \Gamma_\mu \psi_\nu} \, dx^\mu\wedge dx^\nu~, \]
which is related to the supercharge by integrating its real part. By using the three-dimensional fields, $L^\CC_1$ is rewritten as
\begin{multline*}
 L^\CC_1=(\overline{\tilde\eps}\,\gamma_5 T\,\tilde{\psi}_{\underline{i}})\,\theta\wedge\tilde{e}\,{}^{\underline{i}}
+(1-\sqrt{V})(\overline{\tilde{\eps}}\,T\gamma_5\gamma_{\underline{i}}\,\tilde{\chi})\,\theta\wedge \tilde{e}\,{}^{\underline{i}}
\\
+\frac{1}{V}\left(\overline{\tilde{\eps}}\,T\gamma_5\gamma_{\underline{i}}\,\tilde{\psi}_{\underline{j}}
+\overline{\tilde{\eps}}\,\gamma_5 T \gamma_{\underline{i}\,\underline{j}}\,\tilde{\chi}\right)\tilde{e}\,{}^{\underline{i}}\wedge \tilde{e}\,{}^{\underline{j}}~.
\end{multline*}
In the integral over spatial infinity, only the last term is relevant. Its variation under supersymmetry gives, up to terms of order $O(\psi^2)$, 
\[\tilde{L}^\CC_0=
\frac{1}{V}\left(\overline{\tilde{\eps}}\,T\gamma_5\gamma_{\underline{i}}\,\tilde{D}_{\underline{j}}\tilde{\eps}\right)\tilde{e}\,{}^{\underline{i}}\wedge \tilde{e}\,{}^{\underline{j}}+
\frac{1}{V}\left(\overline{\tilde{\eps}}\,\gamma_5 T \gamma_{\underline{i}\,\underline{j}}\gamma^{k}P^I_{k} C_I\,\tilde{\eps}\right)\tilde{e}\,{}^{\underline{i}}\wedge \tilde{e}\,{}^{\underline{j}}~.
\]

For an asymptotically trivial background, $\grpel{V}\rightarrow 1$ and $\tilde{g}_{ij}\rightarrow \eta_{ij}$, the integral of the first summand at spatial infinity is invariant under the compensating $U(1,1)$ transformation. Furthermore the $\su(1,1)$ connection in $\tilde{D}$ drops out from the real part, due to antisymmetry in the inner-product. This term should give the total three-momentum of the space, but because there are no gravitational degrees of freedom in three-dimensions we shall put it to zero. On the other hand the imaginary part of the first term, by letting $1/V\rightarrow 1$, is an exact term so we will ignore it as well. 

In the second summand only the wedge of the Clifford product $\gamma_{\underline{i}\underline{j}}\cdot\gamma^k$ contributes to the integral over $S^2_\infty$, because the coset fields are continuous on any large two-sphere. This is rewritten, again by letting the factor $1/V\rightarrow 1$, as
\[\tilde{L}'^\CC_0= 2\,\left(\overline{\tilde{\eps}}\,T C_I\,\tilde{\eps}\right)\tilde{\hodge} P^I~.\]
By using the representation of the charges, equation \eqref{eq:Pcharges}, the integral of its real part matches the Nester-Witten formula, equation \eqref{eq:EMPoinc},
\begin{equation}\label{eq:RHSofNW}
 \rbs{\tilde{\eps}_0}{T\,\tilde{\eps}_0} M - \rbs{\tilde{\eps}_0}{i\,\tilde{\eps}_0} Q - \rbs{\tilde{\eps}_0}{i\,\gamma_5\,\tilde{\eps}_0} H~.
\end{equation}
On the other hand, the imaginary part of its integral gives the term found in~\cite{Argurio:2008zt}
\begin{equation}\label{eq:theirterm}
 i\, \rbs{\tilde{\eps}_0}{i\,T\gamma_5\,\tilde{\eps}_0} N~.
\end{equation}

Using the generalized Nester-Witten formula we have reduced the electromagnetic superalgebra of charges, that is equation \eqref{eq:EMPoinc}, to three dimensions in the suggestive form
\[ [ \rbs{\tilde\eps_0}{Q}, \rbs{\tilde\eps_0}{Q} ] = - 2\sqrt{2}\rbs{\tilde{\eps}_0}{T\, C_I\,\tilde{\eps}_0}\begin{pmatrix}M+i\, N\\-Q+i\,H\end{pmatrix}^I~,\]
where the index $I$ runs over the coefficients in the basis $\liel{e}_I$. Indeed, if we expand this on the right-hand side the NUT charge drops out as $ \rbs{\tilde{\eps}_0}{T\, C_w\,\tilde{\eps}_0}=0$ and we get the expression in \eqref{eq:RHSofNW}. We also showed that the complexification of the Nester-Witten form gives the NUT charge, see expression \eqref{eq:theirterm}. However, neither the real part nor the imaginary part are covariant under the coset structure. It is easy to covariantize them, e.g. by replacing $\rbs{\tilde{\eps}_0}{T\, C_I\,\tilde{\eps}_0}$ with $\rbs{\tilde{\eps}_0}{\overline{C}_J\, T\, C_I\,\tilde{\eps}_0}$ or similarly that transforms in the adjoint of $\su(1,1)$. The odd-odd bracket that minimally generalizes the electromagnetic superalgebra of charges to a covariant equation is
\begin{equation}\label{eq:covsulieC}
[\rbs{\tilde\eps_0}{Q}, \rbs{\tilde\eps_0}{Q} ] = \liel{Z}_0 \cdot \mathcal{C}\in\mathfrak{k}~. 
\end{equation}
Here $\liel{Z}_0$ is the asymptotic $\su(1,1)$ triplet, which is generated from the square of the asymptotically constant Dirac spinor $\tilde\eps_0$, and it acts on the charges $\mathcal{C}$ in the representation of $\mathfrak{k}$. The right-hand side is the natural action of the asymptotic symmetry algebra $\RR^3\oplus\su(1,1)$, which is generated by
\[  [ {\tilde\eps_0}, {\tilde\eps_0} ] = \tilde{\xi}_0  +\liel{Z}_0 ~,\]
on the charges of the theory. If we expand the right-hand side of \eqref{eq:covsulieC}, we will see all charges, $M$, $N$, $Q$ and $H$, on an equal footing. Note that this bracket is symmetric as it should be and is equivariant under the coset structure. The term $N_{\underline{0}}$ in \cite{Argurio:2008zt} that originates from the real antisymmetric spinor bilinear clearly belongs to a different representation.

For the BPS states, by letting $\eps_0$ satisfy the BPS projection in \eqref{eq:BPSproj}, the right-hand side of \eqref{eq:covsulieC} vanishes, $\liel{Z}_0\cdot\mathcal{C}=0$. Conversely, the vanishing of $\liel{Z}_0\cdot\mathcal{C}$ requires the charges to satisfy the BPS condition and $\liel{Z}_0$ to be the nilpotent element generated by three-dimensional spinors that satisfy the BPS projection. A simple proof of the converse involves a rotation in $U(1,1)$  on $\liel{Z}_0$ and $\mathcal{C}$ so that the components of $\mathcal{C}$ in \eqref{eq:Pcharges} are real. Solving $\liel{Z}_0\cdot\mathcal{C}=0$ like this is easy and a solution requires the saturation of the BPS bound, while it also fixes $\liel{Z}_0$. This also fixes the BPS projection on the Killing spinors, because the constant Dirac spinors decompose into the subspace that generates $\liel{Z}_0$ and a complement that does not.

\section{Discussion}
It has been argued before that the study of NUT-charged spacetime quantities, like the entropy or action, should not be treated in the same class as asymptotically flat space~\cite{Hunter:1998qe,Hawking:1976jb}. Alternatively, one can study their properties at null infinity~\cite{RamaSen:DualMass81,AshtSen:NutFor1982}. This is also the case for asymptotically locally $\AdS$ spaces, see e.g.~\cite{Cai:2006az,Astefanesei:2004kn}. Here, instead, we worked in the timelike reduced theory. The NUT-charged spacetimes have the same three-dimensional asymptotics as flat spacetime. This allows us to study their Killing superalgebra and its extension as embedded in the flat limit solution.

We performed the reduction of the fermions and found how they transform under the coset structure. We were guided in this by the reduction of the four-dimensional supersymmetry transformation of the gravitino. We also related the symmetry variations of the theory's gauge fields to the irreducible three-dimensional spinor bilinear field $\liel{Z}$. After dimensional reduction, the extended Killing superalgebra is described under the coset structure by using a Killing vector $\tilde{\xi}$ and a gauge-covariant constant scalar $\liel{Z}$ in $\su(1,1)$.

Finally, we turned to the Nester-Witten form. We found it is only one component of an irreducible vector in $\mathfrak{k}$. We covariantized it and found its relation to the asymptotic Killing superalgebra extension. The expression is different to the usual odd-odd bracket of the extended Poincar\'e superalgebra of charges in four dimensions, which gives a $U(1)$-invariant scalar. We did not attempt to interpret the three-dimensional equation in terms of Poisson brackets, as in four-dimensions. If this is possible, notwithstanding the absence of a time evolution in the reduced theory, the vector in $\mathfrak{k}$ might generate translations in the coset space. However note that in our formalism supersymmetry does not generate group transformations.

Our viewpoint was that any superalgebra structure we find should be equivariant with respect to the coset structure of the reduced theory. This led us to find the representations of the fermions, which we did by reducing the four-dimensional gravitino variation. The BPS solutions of $\mathcal{N}=2$ were studied also in \cite{Houart:2009ed}, which considers the Kac-Moody extension of the symmetries, but the fermionic sector was not described there. 
In \cite{Argurio:2008zt,Argurio:2008nb} a particular extension of the Poincar\'e superalgebra was proposed, which comes from the complexification of the Nester-Witten form, but in that work the coset structure is not considered. The term they consider is antisymmetric in the real spinor bilinear and here we showed that it belongs to a different representation of $\su(1,1)$ than the terms that were already present in four dimensions, see \eqref{eq:EMPoinc2}. Nevertheless, we were able to show why the inclusion of \cite{Argurio:2008zt} gives the correct NUT charge, expression \eqref{eq:theirterm}. In the reduced theory we also avoided an ill-defined asymptotic integration surface $S_\infty^2$ in four dimensions.

We concentrated on the symmetries of pure $\mathscr{N}=2$ supergravity. This is a simple theory and the results were straightforward. The authors of~\cite{Bossard:2009at} study the BPS states of a big class of theories that were classified in~\cite{Breitenlohner:1987dg}. These are supergravity theories in four dimensions that upon reduction are described by homogeneous sigma models. They suggest a universal structure of the BPS states, based upon the representations of $\spin^*(2\mathscr{N})$. The latter is a non-compact real form of $\mathfrak{so}(2\mathscr{N},\CC)$ and describes the on-shell states of the theories, see also the lorentzian case~\cite{deWit:1992up}. In the $\mathcal{N}=2$ case $\spin^*(4)=\spin(3)\oplus\mathfrak{sl}(2,\RR)$ is the euclidean spin algebra tensored with the $\su(1,1)$ compensating subalgebra.  We would expect that the extended Killing superalgebra for the large class of theories in \cite{Breitenlohner:1987dg,Bossard:2009at} can be described in a similarly simple way. Note that our results (with the exception of \S\ref{sec:StatBPS} and \S\ref{sec:SusyNW}) hold invariably for \emph{generic} stationary backgrounds.

The problem of extending the Killing superalgebra of supergravity backgrounds is by far an open problem. In Einstein-Maxwell theory the electromagnetic extension suffices to describe the BPS stationary solutions. In particular, the Killing spinors square, for each separate charge configuration, to a unique nilpotent element $\liel{Z}$. In principal, one can always define the extension of an asymptotically flat background, or by our results of an asymptotically locally flat background. However, in higher dimensions the structure of a U-duality covariant extension is far richer, see e.g.~\cite{Englert:2007qb}. 

A significant problem remains the Killing superalgebra extension of the maximally supersymmetric plane wave. In $d=11$ the null isometry of the wave obstructs a maximal algebraic construction~\cite{FigueroaO'Farrill:2008ka}. In light of U-duality and the infinite boost of~\cite{Argurio:2008nb}, which gives a NUT-charged Aichelburg-Sexl pp-wave, a solution to this problem that is also related to our results about the stationary case would be very interesting.

\section*{Acknowledgements} The author would like to thank Bin Chen and his group for their warm hospitality in Beijing.

\appendix

\section{A representation of SU(1,2)}\label{app:su12}
In section~\ref{sec:StatBos} we wrote the pullback of the Maurer-Cartan form in terms of the coset scalars, equations  \eqref{eq:Qconn} and \eqref{eq:Pmom}. We used an explicit representation of $\su(1,2)$ for this. A Lie algebraic analysis can be found in~\cite{Houart:2009ed,lrr-2008-1}.

We define the $SU(1,2)$ metric
\[\zeta=\begin{pmatrix} 0&0&-i\\0&1&0\\i&0&0\end{pmatrix}\]
and the Cartan involution $\sigma(g)=\eta^{-1}(g^{-1})^{\dagger}\eta$, where $g\in SU(1,2)$, with
\[\eta=\begin{pmatrix} 1&0&0\\0&-1&0\\0&0&1\end{pmatrix}~.\]

The Borel subalgebra $\mathfrak{b}$ is spanned by
\[\liel{a}=\begin{pmatrix} \frac{1}{2}&0&0\\0&0&0\\0&0&-\frac{1}{2}\end{pmatrix}~,\]
which is in $\mathfrak{k}$, that is $\sigma(\liel{a})=-\eta^{-1}\liel{a}^{\dagger}\eta=\liel{a}$, 
and its negative roots
\[\liel{p}(z)=\begin{pmatrix} 0&0&0\\i\,z&0&0\\0&z^*&0\end{pmatrix}~,\]
for $z\in\CC$, and
\[\liel{n}=\begin{pmatrix} 0&0&0\\0&0&0\\1&0&0\end{pmatrix}~.\]
The non-zero Lie brackets are $[\liel{a},\liel{p}(z)]=-1/2\,\liel{p}(z)$, $[\liel{a},\liel{n}]=-\liel{n}$ and 
$[\liel{p}(1),\liel{p}(i)]=-2\,\liel{n}$.
The non-compact generator of the Cartan subalgebra of $\su(1,2)$ is
\[\liel{h}=\frac{1}{4}\begin{pmatrix}i& 0& 0\\0& -2 i& 0\\0& 0& i\end{pmatrix}~.\]

We use the $\mathfrak{u}(1,1)$ basis: 
\begin{align*}
 \liel{t}_x &= \sqrt{2}/2\,\mathrm{proj}_{\mathfrak{h}}\liel{p}(1)~,
&
\liel{t}_y&=\sqrt{2}/2\,\mathrm{proj}_{\mathfrak{h}}\liel{p}(i)~,
&
\liel{t}_z&=1/2\,\mathrm{proj}_{\mathfrak{h}}\liel{n}-\liel{h}\\\intertext{and}
\liel{u}&=-\,\mathrm{proj}_{\mathfrak{h}}\liel{n}-2/3\liel{h}~.&&&&
\end{align*}
The projection onto $\mathfrak{h}$ is given by $\mathrm{proj}_{\mathfrak{h}}\liel{x}=1/2(\liel{x}+\sigma(\liel{x}))$. We use the orthonormal basis of $\mathfrak{k}$:
\begin{align*}
 \liel{e}_v&=\sqrt{2}\,\mathrm{proj}_{\mathfrak{k}}\liel{a}~,
&
\liel{e}_w&=-\sqrt{2}\,\mathrm{proj}_{\mathfrak{k}}\liel{n}~,\\
\liel{e}_\phi&=\,\mathrm{proj}_{\mathfrak{k}}\liel{p}(1)~,
&
\liel{e}_\magn&=\,\mathrm{proj}_{\mathfrak{k}}\liel{p}(i)~.
\end{align*}
Similarly, the projection onto $\mathfrak{k}$ is $\mathrm{proj}_{\mathfrak{k}}\liel{x}=1/2(\liel{x}-\sigma(\liel{x}))$. This basis transforms under the generators $\mathfrak{u}(1,1)$ as we described in section~\ref{sec:StatBos}. 

With $\grpel{V}$ given by
\[\grpel{V}=\exp{(\log{V}\liel{a})}\exp{(\omega_0 \,\liel{n})}\exp{(\sqrt{2}\,\liel{p}(\phi+i h))}~,\]
the pullback $\liel{P}+\liel{Q}$ is given as in equations \eqref{eq:Qconn} and \eqref{eq:Pmom}. We use this decomposition to find the representation of the fermions under $\mathfrak{u}(1,1)$.

Let us here briefly comment on two more formulations of the bosonic sector of the sigma model. The first is generic and uses the ``metric'' matrix $\grpel{M}=\sigma(\grpel{V})^{-1}\grpel{V}$. The matrix describes a smooth diffeomorphism $G/H\rightarrow G$ and is thus gauge-independent. It transforms as $\grpel{M}\mapsto \sigma(g)\,\grpel{M}\,g$ and the lagrangian can be written as $\Tr(\grpel{M}^{-1}d\grpel{M})$. It has been noted before that this formulation does not allow a description of the fermions. 

The second is specific to the coset space and uses the Kinnersley-Ehlers potentials~\cite{Kinn:Gen1973,Kinnersley:1977pg,*Kinnersley:1977ph,*Kinnersley:1978pz,Mazur:1982db}. The matrix $\eta\, M=-\zeta + w w^\dagger$ defines the Kinnersley vector $w$, which satisfies $w^\dagger\,\zeta\,w=-2$. In our conventions the Kinnersley vector is
\[w=\frac{1}{\sqrt{V}} \begin{pmatrix} i \mathcal{E}^\star \\ \sqrt{2}\Lambda \\ 1 \end{pmatrix}~,\]
and the potentials are $\mathcal{E}=V+i\omega_0-|\Lambda|^2$ and $\Lambda=\phi+i\,\magn$. The transformation $w\mapsto g^\dagger w$ induces a transformation on the potentials. Note, finally, that it is quite easy to find the right action of the Borel subgroup on $\grpel{V}$ in the Borel gauge. It corresponds to gauge shifts and scalings of the Kinnersley-Ehlers potentials. A combination of this and the action of $\mathfrak{u}(1,1)$ on the potentials seems the easiest way to compute the group action.

\section{The supergravity connection of the BPS solutions}\label{app:D}
The calculation of the supercovariant derivative for the BPS solutions in section~\ref{sec:StatBPS} is identical to that in~\cite{Argurio:2008zt}, although here we do not use an explicit gamma matrix representation. For completeness, we give the components of $D_\mu$ in our conventions:
\begin{align*}
D_t&=\partial_t +\frac{1}{2}\frac{r-M}{R^4}\Gamma^{\underline{r}\,\underline{0}}\mathcal{Z}\,R
\left(
e^{(a_m-\beta)\Gamma_5}+i\,e^{(a_q-2\beta)\Gamma_5}\Gamma_{\underline{0}}
\right)
\\
D_r&=\partial_r-\frac{i}{2}\frac{\mathcal{Z}}{R(r-M)}\Gamma^{\underline{0}\,\underline{r}}e^{(a_q-2\beta)\Gamma_5}\Gamma_{\underline{r}}
\\
D_\theta&=\partial_\theta+\frac{1}{2}\Gamma^{\underline{{\theta}}\,\underline{r}}
+\frac{1}{2}\mathcal{Z}\,R^{-1}\Gamma^{\underline{0}\,\underline{\phi}}\Gamma_5
\left(
e^{(a_m-\beta)\Gamma_5}+i\,e^{(a_q-2\beta)\Gamma_5}\Gamma_{\underline{0}}
\right)
\\
D_\phi&=\partial_\phi+\frac{1}{2}\cos\theta\,\Gamma^{\underline{\phi}\, \underline{\theta}}+\frac{1}{2}\sin\theta\,\Gamma^{\underline{\phi}\,\underline{r}}
\\
&+N\cos\theta\,\frac{r-M}{R^3}\mathcal{Z}\,\Gamma^{\underline{r}\underline{0}}
\left(
e^{(a_m-\beta)\Gamma_5}+i\,e^{(a_q-2\beta)\Gamma_5}\Gamma_{\underline{0}}
\right)
\\
&-\frac{1}{2}\mathcal{Z}\, R^{-1}\sin\theta\,\Gamma^{\underline{\phi}\,\underline{ r}}
\left(
e^{(a_m-\beta)\Gamma_5}+i\,e^{(a_q-2\beta)\Gamma_5}\Gamma_{\underline{0}}
\right)~.
\end{align*}
In order to factor the spin connection in the above expressions, we often use the relation $N^2+Mr-N\Gamma_5(r-M)=\mathcal{Z}\,R\,e^{(a_m-\beta)\Gamma_5}$. For the part of the supercovariant derivative that depends on the field-strength we use the latter's Clifford algebra expression, equation \eqref{eq:NUTFinCl}. The $r$-dependence of the Killing spinor is solved immediately by using the BPS projection and the relation
\[\left((\frac{r-M}{R})^{\frac{1}{2}}e^{i\beta/2}\right)'=\frac{1}{2}\frac{R^{-1}\mathcal{Z}}{r-M}e^{(\beta-a_m)i}\left((\frac{r-M}{R})^{\frac{1}{2}}e^{i\beta/2}\right)~,\]
which can be derived from $(Re^{i\beta})'=1$.

\bibliographystyle{utphys}
\bibliography{nutsusy}

\end{document}